%% file: main.tex
\documentclass[10pt,twocolumn,letterpaper]{article}

\usepackage{cvpr}
\usepackage[accsupp]{axessibility}

\input{preamble}

\title{Generalizable Radio-Frequency Radiance Fields for Spatial Spectrum Synthesis}

\author{
Kang Yang \quad Yuning Chen \quad Wan Du\\
University of California, Merced\\
{\tt\small \{kyang73, ychen372, wdu3\}@ucmerced.edu}
}

\begin{document}

\twocolumn[{
\renewcommand\twocolumn[1][]{#1}
\maketitle

\input{./tex/0_Banner}

}]

\input{./tex/0_Abstract}

\input{./tex/1_Introduction}

\input{./tex/2_Motivation}

\input{./tex/3_Relatedwork}

\input{./tex/4_Design}

\input{./tex/5_Evaluation}

\input{./tex/6_Conclusion}

{
    \small
    \bibliographystyle{ieeenat_fullname}

\input{main.bbl}
}

\end{document}

%% file: preamble.tex
\def\ourSystem{\textit{GRaF}\xspace}

\def\ie{\textit{i.e.},\xspace}
\def\eg{\textit{e.g.},\xspace}

\def\nerft{\textit{NeRF$^2$}\xspace}

\usepackage{amssymb}
\usepackage{mathtools}
\usepackage{amsthm}
\usepackage{tikz}
\usepackage{amsmath}
\usepackage{colortbl}
\usepackage{eucal}
\usepackage{bm}
\usepackage{multirow}
\usepackage{enumitem}
\usepackage{xspace}
\usepackage{caption}
\usepackage{booktabs}

\usepackage{graphicx}
\usepackage{subcaption}

\usepackage{threeparttable}

\newcolumntype{C}[1]{>{\centering\arraybackslash}p{#1}}
\newcolumntype{L}[1]{>{\raggedright\arraybackslash}p{#1}}

\newlength{\vslen}
\newtheorem{theorem}{Theorem}

\usepackage{wrapfig}

\usepackage{amsmath}

\definecolor{cvprblue}{rgb}{0.21,0.49,0.74}
\usepackage[pagebackref,breaklinks,colorlinks,allcolors=cvprblue]{hyperref}

%% file: tex/0_Banner.tex
\begin{center}
\captionsetup{hypcap=false}
    \vspace{-8mm}
\includegraphics[width=0.98\textwidth,height=5.3cm]{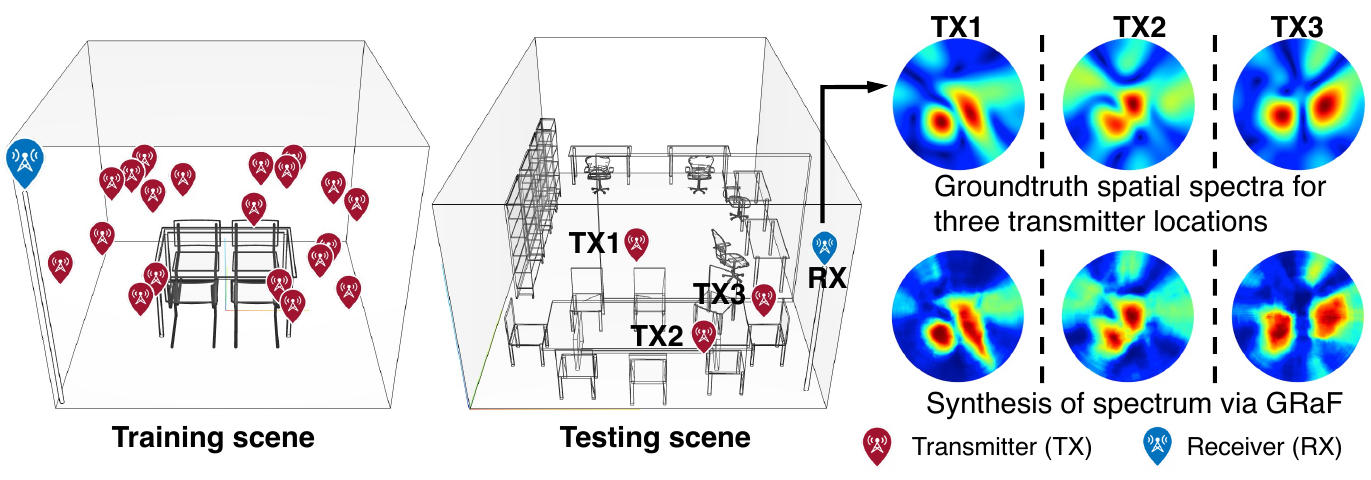}
\captionof{figure}{In the training scene, Radio-Frequency~(RF) signals from each transmitter are measured across all surrounding directions by the receiver to form a spatial spectrum. Trained on this scene, \ourSystem synthesizes spectra for arbitrary transmitter locations in unseen scenes.}
    \label{fig_overall}
\end{center}

%% file: tex/0_Abstract.tex
\begin{abstract}

\vspace{-0.1in}
\noindent
We present \ourSystem, Generalizable Radio-Frequency~(RF) Radiance Fields, a framework that models RF signal propagation to synthesize spatial spectra at arbitrary transmitter or receiver locations, where each spectrum measures signal power across all surrounding directions at the receiver. 
Unlike state-of-the-art methods that adapt vanilla Neural Radiance Fields~(NeRF) to the RF domain with scene-specific training, \ourSystem generalizes across scenes to synthesize spectra. 
To enable this, we prove an interpolation theory in the RF domain: the spatial spectrum from a transmitter can be approximated using spectra from geographically proximate transmitters.
Building on this theory, \ourSystem comprises two components: 
\textit{(i)}~a geometry-aware Transformer encoder that captures spatial correlations from neighboring transmitters to learn a scene-independent latent RF radiance field, and 
\textit{(ii)}~a neural ray tracing algorithm that estimates spectrum reception at the receiver.
Experimental results demonstrate that \ourSystem outperforms existing methods on single-scene benchmarks and achieves state-of-the-art performance on unseen scene layouts.
Our source code is publicly available at \href{https://github.com/kangyang73/GRaF}{https://github.com/kangyang73/GRaF}.

\end{abstract}

%% file: tex/1_Introduction.tex
\vspace{-0.2in}
\section{Introduction}\label{sec_introduction}

The evolution of wireless communication systems, such as WiFi and Sixth Generation~(6G) cellular networks, has enabled transformative applications in domains such as smart cities and smart healthcare~\cite{gadre2020frequency, tong2023citywide, li2025exploring, ullah2024survey}.
These networks function as both communication backbones and sensing platforms. 
Reliable communication requires meticulous network planning to mitigate issues such as dead spots and outages~\cite{ahamed20215g}, while sensing applications, \eg WiFi-based localization~\cite{ayyalasomayajula2020deep, savvides2001dynamic}, depend on large volumes of high-quality data to train deep neural networks~(DNNs).
Site surveys~\cite{kar2018site, yang2024orchloc, yang2025generative} are conducted to collect data for communication and sensing.
However, conducting these surveys demands dense measurements at numerous locations, which is both time-consuming and labor-intensive~\cite{site_survey_cisco, kovari2025synergizing}.

An alternative is to synthesize Radio-Frequency~(RF) datasets via propagation modeling, generating spatial spectra that capture the received power from all directions around the receiver~(§\ref{sec_spatial_spectrum}).
Given transmitter and receiver positions, RF propagation modeling estimates the received spectrum while accounting for reflection, diffraction, and scattering~\cite{1451581, na2022huygens}.
In free space, Maxwell’s equations~\cite{maxwell1873treatise} enable accurate signal computation~\cite{wong1984conductivity}; however, in complex real-world environments, directly solving them becomes intractable~\cite{zhao2000integral}.
To approximate these scenarios, ray tracing simulations~\cite{yun2015ray, egea2021opal, matlab_indoor_simulation, guo2024rad} model propagation paths using Computer-Aided Design~(CAD) scene models~\cite{liu2024wireless, modesto2025accelerating}.
However, obtaining accurate CAD models and the high computational cost of ray tracing make such simulations both impractical and unreliable in practice.

Recently, \nerft~\cite{zhao2023nerf} and NeWRF~\cite{lunewrf} have adapted Neural Radiance Fields~(NeRF)~\cite{mildenhall2021nerf, tancik2022block, liu2020neural, martin2021nerf}, originally designed for novel view synthesis in the optical domain, to model RF signal propagation and synthesize spectrum.
These methods achieve state-of-the-art performance in RF data synthesis by learning neural radiance fields tailored to RF signals.
However, like vanilla NeRF, they tend to overfit the scenes they are trained on, requiring time-intensive retraining for each new scene~\cite{mildenhall2021nerf, wang2021ibrnet, liu2022neural}.

This paper introduces \ourSystem, which enhances RF spatial spectrum synthesis within a scene and generalizes to unseen scenes.
As illustrated in Figure~\ref{fig_overall}, spectra collected from multiple transmitter locations in a training scene are used to train \ourSystem.
After training, \ourSystem can synthesize high-quality spatial spectra for arbitrary locations in new scenes.
This capability is supported by our \emph{Interpolation~Theorem~\ref{theory_interpolation}}~(§\ref{sec_interpolation}), which establishes that the spectrum at any transmitter location can be approximated by interpolating the spectra of geographically proximate transmitters.

Building on this theory, \ourSystem learns a latent RF radiance field from the spectra of neighboring transmitters, enabling spatial spectrum synthesis beyond the training scene. 
Constructing such a field requires a fundamentally different approach from optical radiance fields, due to the distinct propagation characteristics of RF signals. 
Unlike visible light, 
RF signals have centimeter-scale wavelengths and interact with obstacles through more complex mechanisms, including absorption, reflection, diffraction, and scattering~\cite{1451581, na2022huygens}. 
To account for these interactions, we represent voxel attributes as latent vectors in a learned radiance field, constructed using a geometry-aware Transformer encoder with cross-attention. 
This encoder processes the spatial spectra of neighboring transmitters along with their geometric relationships to the target transmitter location.

Furthermore, we propose a neural ray tracing algorithm that operates on the latent RF radiance field to estimate the spatial spectrum. 
This algorithm traces rays outward from the receiver, sampling voxels along each path to account for interactions within the scene.
At each voxel, it predicts complex-valued contributions that capture both amplitude and phase variations due to RF propagation.
These contributions are then aggregated along each ray, incorporating free-space path loss and phase shifts induced by propagation delay, to compute the received spatial spectrum.

The key contribution of this work is a new paradigm for generalizable RF spatial spectrum synthesis, grounded in an interpolation theory that estimates the spectrum at a target location from the spatial spectra of nearby transmitters. 
\ourSystem replaces scene-specific NeRF-style approaches with a geometry-aware Transformer that encodes spatial correlations from neighboring transmitters into a latent RF radiance field, followed by a neural ray tracing algorithm that aggregates these features to compute the spatial spectrum.
We demonstrate that \ourSystem achieves state-of-the-art performance in both single-scene and cross-scene settings, and further enhances downstream localization performance.

%% file: tex/2_Motivation.tex
\section{Preliminary}\label{sec_motivation}

\begin{figure}[t]
\centering
\begin{subfigure}{.165\textwidth}
  \centering
  \includegraphics[width=\linewidth]{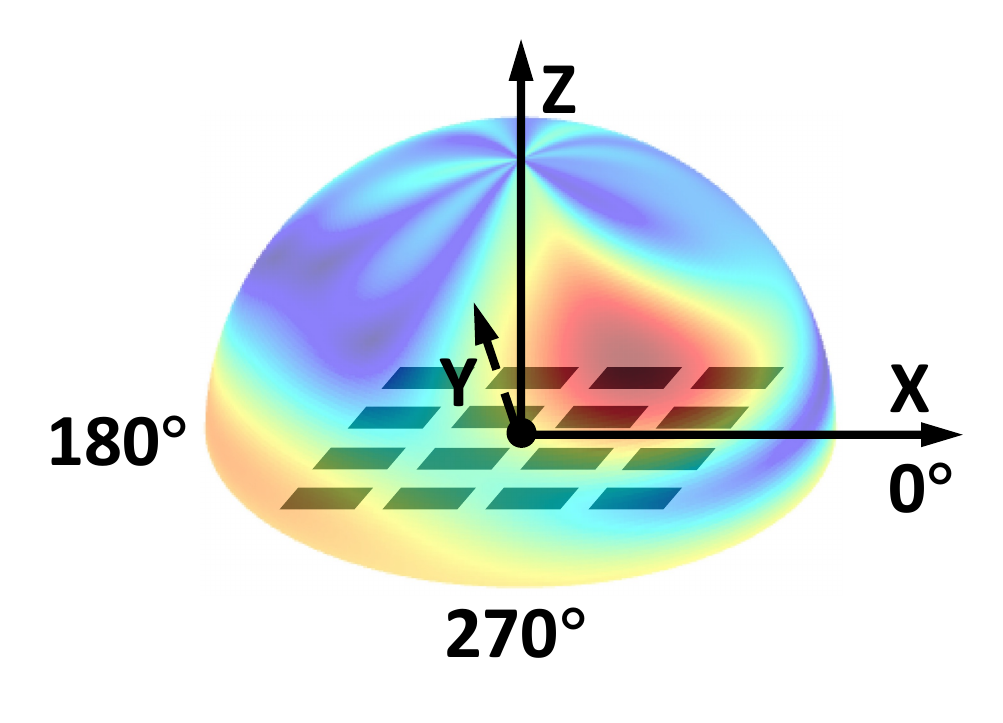}
  \caption{3D spherical view}
  \label{fig_spectrum_a}
\end{subfigure}
\hspace{2em}
\begin{subfigure}{.145\textwidth}
  \centering
  \includegraphics[width=\linewidth]{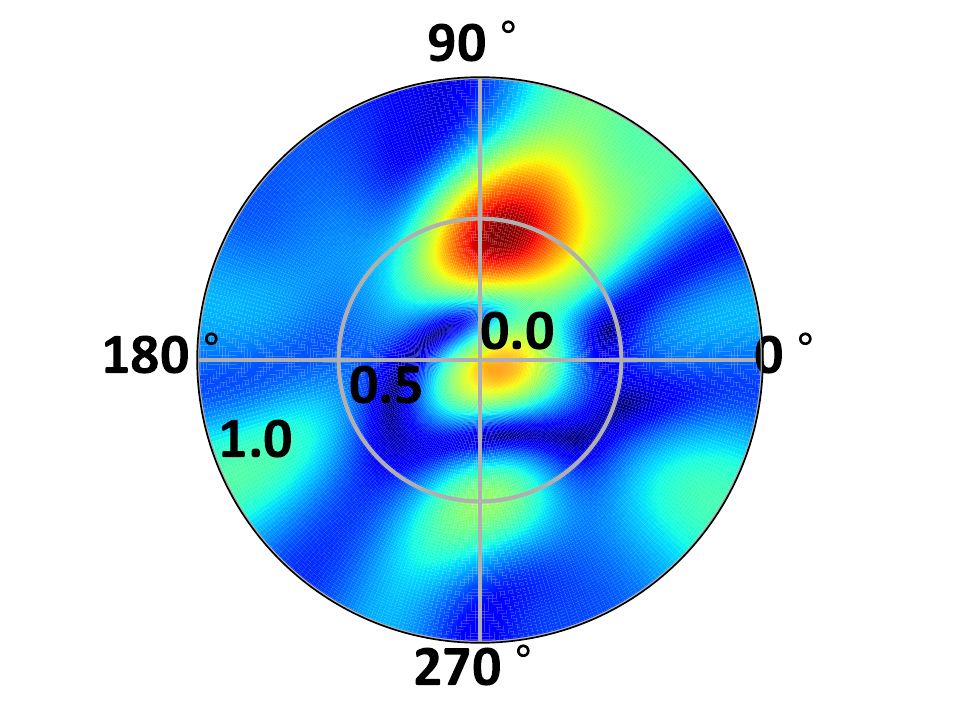}
  \caption{2D projection view}
  \label{fig_spectrum_b}
\end{subfigure}
\caption{Illustration of the spatial spectrum in 3D and 2D views.}
\label{fig_spectrum}
\end{figure}


\subsection{Spatial Spectrum}\label{sec_spatial_spectrum}

The spatial spectrum characterizes how received RF signal power is distributed across 3D directions~(azimuth and elevation) around the receiver when the transmitter emits RF signals.
We first present the wireless channel model and then describe how the spatial spectrum is computed.

\noindent
\textbf{Wireless Channel Model.}
A wireless communication system consists of a transmitter and a receiver. 
The transmitted signal can be represented as a complex number~$x = A e^{j\psi}$, where~$A$ is the amplitude and~$\psi$ is the phase. 
As the signal propagates through the wireless channel, its amplitude is attenuated by~$A_{\text{att}}$, and its phase is shifted by~$\Delta \psi$. 
The signal received at a single-antenna receiver is given by:
\begin{equation}
y = x \cdot A_{\text{att}} e^{j\Delta \psi} = A \cdot A_{\text{att}} e^{j(\psi + \Delta \psi)}.
\end{equation}

For a receiver configured as a~$M \times N$ antenna array, the received signal is represented as a matrix~$\mathbf{Y} = \big[y_{m,n}\big]_{m,n=0}^{M-1, N-1}$, where~$y_{m,n}$ denotes the signal received at the~$(m, n)$-th array element.
Each array element $y_{m,n}$ is computed as follows:
\begin{equation}
y_{m,n} = A \cdot A_{\text{att}} e^{j(\psi + \Delta \psi + \Delta \sigma_{m,n})},
\end{equation}
where~$A_{\text{att}} e^{j\Delta \psi}$ is the attenuation and phase shift introduced by the channel.
The term~$\Delta \sigma_{m,n}$ is the geometric phase shift at the~$(m, n)$-th element, caused by its position relative to the array center at~$(0, 0, 0)$.
It is determined by the element position and the angles of arrival:
\begin{equation}
\label{eqn_geomtry}
\Delta \sigma_{m,n} = \frac{2\pi}{\lambda} \left( m d \sin\beta \cos\alpha + n d \sin\beta \sin\alpha \right),
\end{equation}
where $\lambda$ is the signal wavelength, $d = \frac{\lambda}{2}$ is the element spacing, and $\alpha$ and $\beta$ are the azimuth and elevation angles of the incoming signal, respectively.

\noindent
\textbf{Spatial Spectrum Computation.}
It is computed from the received signal matrix~$\mathbf{Y}$ to estimate the power distribution over azimuth~$\alpha$ and elevation~$\beta$ angles. 
For a given direction~$(\alpha, \beta)$, the received signal power is given by:
\begin{equation}
\mathbf{SS}(\alpha, \beta) = \left| \sum_{m=0}^{M-1} \sum_{n=0}^{N-1} y_{m,n} e^{-j \Delta \sigma_{m,n}} \right|^2,
\end{equation}
where~$\Delta \sigma_{m,n}$ is the theoretical geometric phase shift defined in Equation~(\ref{eqn_geomtry}). 
With one-degree resolution and only the upper hemisphere considered, the spatial spectrum $\mathbf{SS}$ is represented as a matrix of size $(N_a, N_e)$:
\(
\mathbf{SS} = \big[ SS_{p,q} \big]_{p=0,q=0}^{N_a-1,N_e-1},
\)
where $N_a = 360$~(azimuth) and $N_e = 90$~(elevation).
Figure~\ref{fig_spectrum_a} shows the 3D spatial spectrum, while Figure~\ref{fig_spectrum_b} depicts the corresponding 2D projection onto the~X-Y plane, where the radial distance is~$\cos(\beta)$.

\subsection{Vanilla NeRF for Spatial Spectrum Synthesis}
\label{sec_nerf2_intro}

Vanilla NeRF has recently been adapted for RF spatial spectrum synthesis~\cite{zhao2023nerf, lunewrf}. 
It models a 3D scene using a Multi-Layer Perceptron~(MLP) with 8-dimensional inputs:
\begin{equation}
\label{eqn_nerf2_voxel}
\mathcal{F}_{\Theta} : \left( \{x, y, z\}, \alpha, \beta, \{x_{tx}, y_{tx}, z_{tx}\} \right) \to \left( \delta, \xi \right),
\end{equation}
where~$\{x, y, z\}$ are the coordinates of a voxel, and~$(\alpha, \beta)$ denote the azimuth and elevation angles defining the view direction of a ray traced from the receiver. 
The transmitter coordinates~$\{x_{tx}, y_{tx}, z_{tx}\}$ are also included, as each voxel’s properties are influenced by the transmitter’s location. 
The outputs~$(\delta, \xi)$ represent the voxel’s RF characteristics, where~$\delta$ is the attenuation and~$\xi$ is the signal emission, with each voxel treated as a secondary RF signal source~\cite{born2013principles}.

To generate spatial spectrum, \nerft~\cite{zhao2023nerf} performs ray tracing in each direction of the spectrum. 
Discrete sample voxels~$\{V_1, \ldots, V_S\}$ are taken along each ray~$\mathbf{r}$, and the~MLP is queried with their coordinates to predict~$\delta$ and~$\epsilon$. 
The result for each ray,~$\mathbf{SS}(\mathbf{r})$, is calculated as:
\vspace{-5pt}
\begin{equation}
\label{eqn_tracing_rf}
\mathbf{SS}\left(\mathbf{r}\right) = \sum_{i=1}^{S} \exp \left( \sum_{j=1}^{i-1} \delta_j \right) \xi_i,
\end{equation}
where RF signals re-emitted by voxels along the ray are aggregated, with each voxel acting as a source whose contribution is attenuated by intervening voxels.

%% file: tex/3_Relatedwork.tex
\section{Related Work}\label{sec_relatedWork}

\textbf{Generalization in Optical NeRFs.}  
Optical NeRFs have revolutionized novel view synthesis by learning scene representations from images.
However, vanilla NeRF requires extensive per-scene training, limiting its generalization to new scenes~\cite{mildenhall2021nerf, wang2021ibrnet, tancik2022block}. 
Generalizable NeRFs overcome this limitation by adapting to unseen scenes without retraining~\cite{liu2022neural, chen2021mvsnerf, trevithick2020grf, 10377995, chou2024gsnerf, tian2023mononerf, yu2021pixelnerf, chen2023gm, zhu2024caesarnerf, bergman2022generative, yang2023freenerf}. 
For example, MVSNeRF integrates multi-view stereo with neural rendering to reconstruct radiance fields from a few input views~\cite{chen2021mvsnerf}, and WaveNeRF employs wavelet-based representations for generalization~\cite{10377995}. 
GSNeRF incorporates semantics to generate novel views and semantic maps for unseen scenes~\cite{chou2024gsnerf}.
These methods cannot be directly applied to RF signals due to fundamental differences in wavelength and propagation behaviors.
This work extends generalizable NeRFs to~RF domain for spatial spectrum synthesis.

\noindent
\textbf{RF Spatial Spectrum Synthesis.}
Several recent works have explored NeRF-style methods in the RF domain~\cite{zhao2023nerf, orekondy2022winert}.
For example, \nerft~\cite{zhao2023nerf} trains an MLP to compute voxel attributes and uses Equation~(\ref{eqn_tracing_rf}) for ray tracing to compute the spatial spectrum. 
However, it requires training a separate model for each scene, limiting scalability and hindering generalization to new scenes.
NeWRF~\cite{lunewrf} leverages direction-of-arrival~(DoA) priors to reduce ray sampling and improve efficiency, but it requires specialized antenna arrays for DoA measurements, which are often impractical.
WiNeRT~\cite{orekondy2022winert} employs differentiable ray tracing with CAD-based scene geometry, but accurate~CAD models are difficult to obtain in practice.
RFScape~\cite{chen2025radio} models RF propagation using neural Signed Distance Function~(SDF)–based scene geometry, but it requires known object shapes and layouts, limiting practicality.
In contrast, \ourSystem requires no scene-specific training, no DoA measurements, and no prior scene models, making it a practical solution for RF spatial spectrum synthesis.
GeRaF~\cite{lu2025geraf} reconstructs 3D object geometry from near-field RF scans, whereas our work models scene-level RF spectrum synthesis and generalizes across different scenes.

3DGS~\cite{kerbl20233d, yu2024mip, yang2025scalable}-based spectrum synthesis methods~\cite{zhang2024rf, wen2024wrf, yanggsrf, zhang2025rf, li2025wideband} improve the training and inference times of~\nerft~\cite{zhao2023nerf} within a single scene. 
In contrast, our work focuses on achieving spatial generalization across scenes.

%% file: tex/4_Design.tex
\section{Methodology}

\begin{figure*}[t]
\centering
{\includegraphics[width=\textwidth,height=5.5cm]{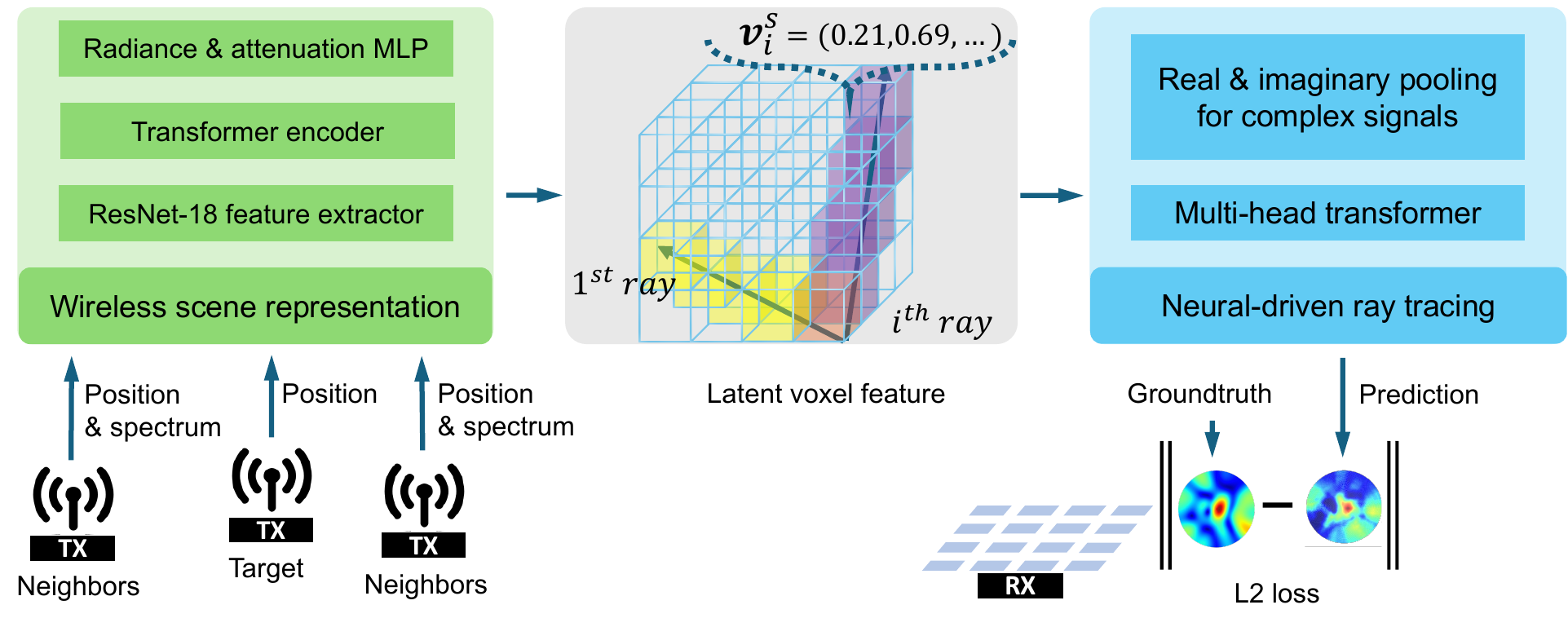}}
\caption{Architecture of \ourSystem. Each voxel is represented by a feature $\mathbf{v}^s_i$, where $i$ indexes the $M$ rays, and $s$ denotes the voxel's position along the $i$-th ray. A neural-driven ray tracing algorithm computes the received signal power for each ray (TX: transmitter, RX: receiver).}
	\label{fig_workflow}
  \vspace{-0.2in}
\end{figure*}

Given a set of training scenes~$\{1,\dots,K\}$, each containing spatial spectra and transmitter positions 
\(
\mathcal{S}^{(k)} = \left\{ \big( \mathbf{SS}_i^{(k)}, \mathbf{P}_i^{(k)} \big) \right\}_{i=1}^{N_k},
\)
where~$\mathbf{SS}_i^{(k)} \in \mathbb{R}^{360 \times 90}$ denotes the spatial spectrum and~$\mathbf{P}_i^{(k)} \in \mathbb{R}^3$ is the transmitter position, the objective is to learn a model $\mathcal{F}_\Theta$ that synthesizes the spectrum $\mathbf{SS}^{(j)}_{\text{target}}$ for a transmitter at position $\mathbf{P}^{(j)}_{\text{target}}$ in any scene~$j$, using its~$L$ nearest neighboring transmitters $\mathcal{N}_L^{(j)} \subset \mathcal{S}^{(j)}$.
Justification is given in §F.1 of the \textit{Supplementary Materials}.
This objective can be formulated as follows:
\begin{equation}
\Theta^* = \arg\max_{\Theta} \prod_{k=1}^K \prod_{i=1}^{N_k} 
p\!\left( \mathbf{SS}_i^{(k)} \,\middle|\, \mathcal{N}_L^{(k)}\!\left( \mathbf{P}_i^{(k)} \right), \mathbf{P}_i^{(k)}, \Theta \right),
\label{eqn_objective}
\end{equation}
where~$\mathcal{N}_L^{(k)}\!\left( \mathbf{P}_i^{(k)} \right)$ denotes the $L$ nearest transmitters to~$\mathbf{P}_i^{(k)}$ in $\mathcal{S}^{(k)}$.
At inference time, for any scene $j$ (seen or unseen during training), given a target transmitter position~$\mathbf{P}_{\text{target}}^{(j)}$ and its neighbors $\mathcal{N}_L^{(j)} \subset \mathcal{S}^{(j)}$, the learned model synthesizes the corresponding spatial spectrum as:
\begin{equation}
\mathbf{SS}_{\text{target}}^{(j)} = 
\mathcal{F}_{\Theta^*}\!\left( \mathcal{N}_L^{(j)}\!\left( \mathbf{P}_{\text{target}}^{(j)} \right), \mathbf{P}_{\text{target}}^{(j)} \right).
\end{equation}

\subsection{RF Spatial Spectrum Interpolation Theory}\label{sec_interpolation}

\begin{theorem}[Spectrum Interpolation]
\label{theory_interpolation}
Let $\mathbf{SS}(\mathbf{P})$ denote the spatial spectrum of a transmitter at position $\mathbf{P} \in \mathbb{R}^3$, and let $\mathcal{N}_{L}(\mathbf{P}) = \{\mathbf{P}_i\}_{i=1}^{L}$ be its $L$ nearest neighbors with spectra $\mathbf{SS}_i$. 
Then the spectrum at $\mathbf{P}$ admits the approximation
$\mathbf{SS}(\mathbf{P}) \approx \sum_{i=1}^{L} w_i\, \mathbf{SS}_i$,
for barycentric weights $\{w_i\}$ determined by the local transmitter geometry.  
The interpolation error satisfies $\epsilon \le K\,\delta^{2}$, where $\delta = \max_i \|\mathbf{P}-\mathbf{P}_i\|$ denotes the neighborhood radius and $K$ characterizes the environment curvature.
The full theorem and proof are provided in §A of the \textit{Supplementary Materials}.
\end{theorem}

\subsection{Model Overview}\label{sec:model_overview}

As illustrated in Figure~\ref{fig_workflow}, \ourSystem comprises two key components:~(i) a latent RF radiance field learned by a geometry-aware Transformer that captures spatial correlations from neighboring transmitters, and (ii) a neural ray tracing algorithm that aggregates latent voxel features along directional rays to compute the spatial spectrum.

Central to this process is a latent variable~$\mathbf{Z}$~\cite{meng2021gnerf} that encapsulates the scene’s RF propagation characteristics, such as path loss and multipath effects. 
In particular, $\mathbf{Z}$ represents the interpolation weights from Theorem~\ref{theory_interpolation}, refined through non-linear transformations to capture propagation behaviors beyond simple linear combinations. 
Accordingly, the likelihood of the spatial spectrum $\mathbf{SS}$ given the $L$ nearest neighbors $\mathcal{N}_L$ and transmitter position $\mathbf{P}$ is given by:
\begin{equation}
p\left( \mathbf{SS} \,\middle|\, \mathcal{N}_L, \mathbf{P} \right)
= \int p\left( \mathbf{SS} \,\middle|\, \mathbf{Z}, \mathbf{P} \right)\, p\left( \mathbf{Z} \,\middle|\, \mathcal{N}_L \right)\, d\mathbf{Z},
\end{equation}
where $\mathcal{N}_L = \mathcal{N}_L(\mathbf{P})$ denotes the $L$ nearest transmitters to $\mathbf{P}$.
Since $\mathbf{Z}$ encodes global RF propagation properties~(\eg materials and geometry), the signal power along each ray depends only on $\mathbf{Z}$. 
The neural ray tracing algorithm applies the interpolation theory directionally by deriving ray-specific weights from $\mathbf{Z}$ to estimate the spectrum for each direction. 
Thus, the likelihood decomposes across rays as:
\begin{equation}
p\left( \mathbf{SS} \,\middle|\, \mathcal{N}_L, \mathbf{P} \right)
= \int \prod_{r=1}^{Q} p\left( \mathbf{SS}(r) \,\middle|\, \mathbf{Z}, \mathbf{P} \right)\, p\left( \mathbf{Z} \,\middle|\, \mathcal{N}_L \right)\, d\mathbf{Z},
\end{equation}
where $Q = N_a \times N_e$ denotes the total number of rays discretized over azimuth $(\alpha)$ and elevation $(\beta)$ angles.
For example, $Q = 360 \times 90$ at one-degree resolution, covering the upper hemisphere centered at the receiver.

\subsection{Latent RF Radiance Field}\label{sec_design_latent_rf}

The latent RF radiance field computes the latent variable~$\mathbf{Z}$, which encodes the scene’s RF propagation characteristics, including path loss, shadowing, and multipath interactions~(\eg reflection and diffraction). 
This contextual representation enables \ourSystem to generalize spectrum synthesis across diverse scenes by providing a unified encoding of RF propagation behavior. 
The latent variable $\mathbf{Z}$ is obtained using a parameterized function $\mathcal{T}_\Psi$ that takes as input the $L$ nearest neighbors $\mathcal{N}_L$ and the transmitter position $\mathbf{P} \in \mathbb{R}^3$:
\begin{equation}
\mathbf{Z} = \mathcal{T}_\Psi\left(\mathcal{N}_L, \mathbf{P}\right).
\end{equation}

The architecture of~$\mathcal{T}_\Psi$ is designed to effectively integrate both spectral and geometric information from $\mathcal{N}_L$. 
It begins with a ResNet-18 feature extractor that processes each neighbor’s spatial spectrum $\mathbf{SS}_i \in \mathcal{N}_L$, producing a compact feature vector that captures high-level patterns such as directional power distribution and signal strength variations.
In parallel, the relative positions $(\mathbf{P}_i - \mathbf{P})$~(for each neighbor $\mathbf{P}_i \in \mathcal{N}_L$) are encoded using positional embeddings to preserve the spatial relationships between the target transmitter and its neighbors. 
These spectral features and geometric embeddings capture both the RF signal characteristics and the geometric context of the scene.

The joint spectral–geometric representations are then processed by a geometry-aware Transformer encoder equipped with cross-attention layers. 
The cross-attention mechanism dynamically emphasizes the most influential neighboring transmitters by weighting their spectral and geometric features, effectively learning the interpolation weights described in Theorem~\ref{theory_interpolation} while capturing non-linear propagation effects such as interference, diffraction, and multipath scattering. 
The output of the Transformer encoder is subsequently refined by two MLPs, which map the intermediate representation to the final latent variable~$\mathbf{Z} \in \mathbb{R}^d$. 
The resulting~$\mathbf{Z}$ serves as a compact contextual encoding that summarizes the scene’s RF propagation behavior and the spatial relationships among neighboring transmitters, enabling \ourSystem to synthesize spatial spectra.

\subsection{Neural Ray Tracing Algorithm}\label{sec_design_ray}

The neural ray tracing algorithm synthesizes the spatial spectrum $\mathbf{SS}_\Theta$ by leveraging the contextual latent variable $\mathbf{Z}$. 
It traces directional rays from the receiver and aggregates latent features along each ray to estimate the signal power in that direction, producing the complete spatial spectrum.

For each ray~$r$ with direction~$(\alpha, \beta)$, we sample~$S$ points~$\{\mathbf{x}_s\}_{s=1}^{S}$ along its path, evenly spaced to cover the spatial extent of the scene. 
At each sampled point~$\mathbf{x}_s$, a voxel-specific feature $\mathbf{v}_s \in \mathbb{R}^d$ is computed by combining the latent variable $\mathbf{Z}$, the transmitter position $\mathbf{P} \in \mathbb{R}^3$, and positional encodings of $\mathbf{x}_s$ and the ray direction $(\alpha, \beta)$:
\begin{equation}
\mathbf{v}_s = \mathrm{MLP}\!\left(\mathbf{Z},\; \mathrm{PosEnc}\!\left(\mathbf{x}_s, (\alpha, \beta), \mathbf{P}\right)\right),
\end{equation}
where $\mathrm{MLP}$ is a multi-layer perceptron and $\mathrm{PosEnc}$ denotes the positional encoding function. 
The resulting features are then processed by an RF-related function $\mathcal{R}_\Phi$ to estimate the signal power for the ray:
\begin{equation}
\hat{\mathbf{SS}}_\Theta(r) = \mathcal{R}_\Phi\left(\{\mathbf{v}_s\}_{s=1}^{S}\right).
\end{equation}

To incorporate RF-specific characteristics, $\mathcal{R}_\Phi$ models the radiated complex signal and the complex-valued attenuation at each sampled point $\mathbf{x}_s$ along the ray. 
Inspired by the wireless channel model, the radiated signal $s(\mathbf{x}_s, \alpha, \beta)$ and attenuation $a(\mathbf{x}_s, \alpha, \beta)$ are computed as:
\begin{equation}
\begin{aligned}
s(\mathbf{x}_s, \alpha, \beta) 
&= I_s(\mathbf{x}_s, \alpha, \beta) + j\, Q_s(\mathbf{x}_s, \alpha, \beta) \\
&= \mathrm{MLP}_s(\mathbf{v}_s), \\
a(\mathbf{x}_s, \alpha, \beta) 
&= A_s(\mathbf{x}_s, \alpha, \beta) + j\, B_s(\mathbf{x}_s, \alpha, \beta) \\
&= \mathrm{MLP}_a(\mathbf{v}_s),
\end{aligned}
\end{equation}
where $\mathrm{MLP}_s$ and $\mathrm{MLP}_a$ map the feature $\mathbf{v}_s$ to the complex signal and complex attenuation, respectively. 
The attenuation term $a(\mathbf{x}_s, \alpha, \beta)$ encodes both amplitude reduction and phase shift introduced at $\mathbf{x}_s$. 
The received ray signal~$y_r$ is then obtained by aggregating the contributions from all sampled points along the ray. 
A detailed derivation and proof are provided in~§B of the \textit{Supplementary Materials}.
\begin{equation}
y_r = \sum_{s=1}^S \left( \prod_{j=1}^{s-1} a\left(\mathbf{x}_j, \alpha, \beta\right) \right) s\left(\mathbf{x}_s, \alpha, \beta\right) \cdot \frac{\lambda}{4 \pi d_s} e^{-j \frac{2 \pi f d_s}{c}},
\end{equation}
this aggregation process can be modeled using a Transformer by mapping voxel-wise features~$(\delta, \xi)$ into token features, where the attention scores naturally represent cumulative attenuation along the ray. 
Finally, the predicted spatial spectrum value for ray $r$ is obtained as the power of its received signal:
\(
\hat{\mathbf{SS}}_\Theta(r) = |y_r|^2.
\)

This procedure is applied to all rays, which discretize the spectrum at one-degree resolution over azimuth and elevation angles, resulting in the complete spatial spectrum $\mathbf{SS}_\Theta$.

\noindent
\textbf{Loss Function.}
With the spectrum $\mathbf{SS}_\Theta$ computed, the overall parameters $\Theta$, which include $\Psi$ (from $\mathcal{T}_\Psi$) and $\Phi$~(from $\mathcal{R}_\Phi$), are optimized to maximize the log-likelihood of the observed spatial spectra, as defined in Equation~(\ref{eqn_objective}):
{\small
\begin{equation}
\begin{aligned}
\Theta^*
&= \arg\max_{\Theta} \log p\left(\mathbf{SS}_\Theta \mid \mathcal{N}_L, \mathbf{P}\right) \\
&= \arg\max_{\Theta} \log \int \prod_{r=1}^{Q} p\left(\mathbf{SS}(r) \mid \mathbf{Z}, \mathbf{P}\right)\,
p\left(\mathbf{Z} \mid \mathcal{N}_L\right)\, d\mathbf{Z}.
\end{aligned}
\end{equation}
}
Direct computation of this integral is intractable due to the high dimensionality of $\mathbf{Z}$. 
Because $\mathbf{Z}$ is deterministically produced by the latent RF radiance field, the optimization reduces to a supervised reconstruction problem.
The derivation is provided in~§C of the {\normalfont\textit{Supplementary Materials}}.
\begin{equation}
\Theta^* = \arg\min_{\Theta} \sum_{r=1}^{Q} \left\| \mathbf{SS}(r) - \hat{\mathbf{SS}}_\Theta(r) \right\|^2.
\end{equation}

%% file: tex/5_Evaluation.tex
\section{Experiments}\label{sec_evaluation}

\begin{figure}[tp]
\centering
\begin{subfigure}{.23\textwidth}
  \centering
  \includegraphics[width=\linewidth]{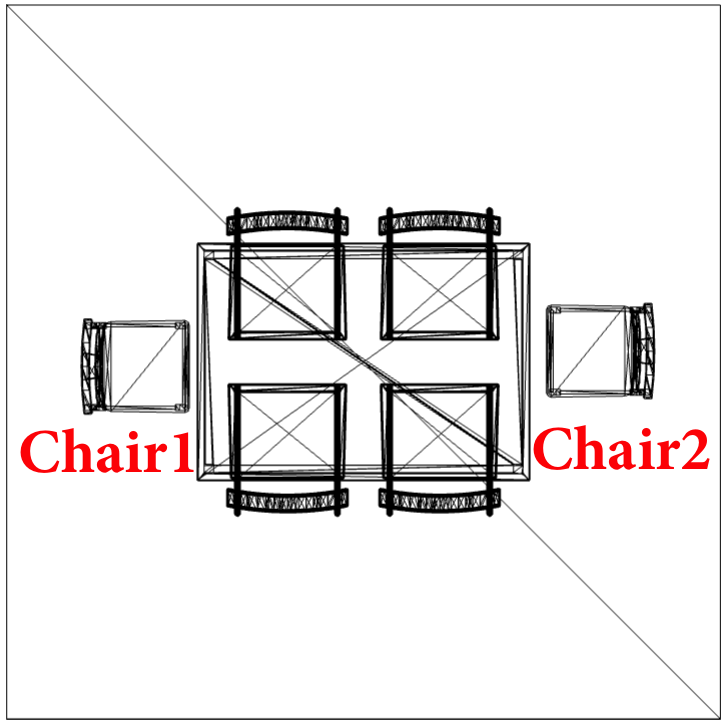}
  \caption{Conference room}
  \label{fig_exp_3d_a}
\end{subfigure}
\hspace{1em}
\begin{subfigure}{.145\textwidth}
  \centering
  \includegraphics[width=\linewidth]{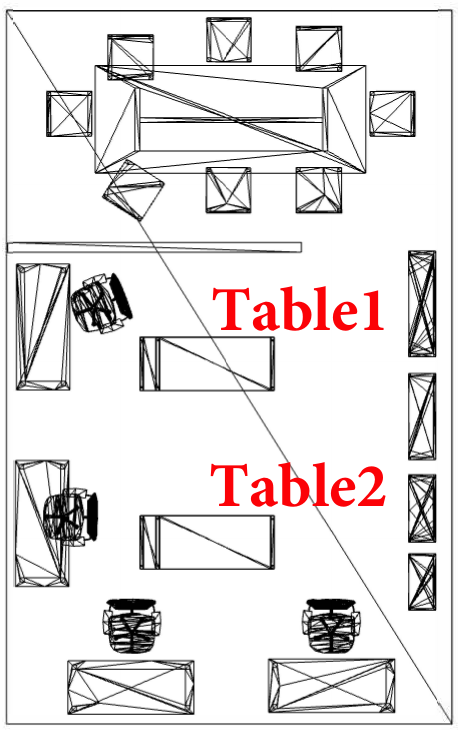}
  \caption{Office room}
  \label{fig_exp_3d_b}
\end{subfigure}
\caption{Top-view visualization of two room layouts.}
\label{fig_exp_3d}
\end{figure}

Implementation details are provided in~§D of the \textit{Supplementary Materials}.
We first evaluate \ourSystem in a single-scene setting, and then assess its cross-scene performance.

\subsection{Experimental Setup}\label{sec_experimental_setting}

\textbf{\underline{\textbullet~Datasets}}~\textsc{(i)~RFID Dataset:}
Introduced by \nerft~\cite{zhao2023nerf}, RF Identification~(RFID) operates in 915\,MHz, with a receiver positioned at a fixed location and equipped with a~$4 \times 4$ antenna array. 
The dataset consists of~6,123 transmitter locations, each associated with a spectrum.

\textsc{(ii)~MATLAB Dataset:}
We consider two representative indoor layouts: a conference room and an office room. 
Their geometries are illustrated in Figure~\ref{fig_exp_3d}, with dimensions of~\(9.8\,\text{ft} \times 9.8\,\text{ft} \times 8.2\,\text{ft}\) and~\(26.2\,\text{ft} \times 16.4\,\text{ft} \times 9.8\,\text{ft}\), respectively.
Each layout is represented by a CAD model and further diversified into multiple scene versions by modifying object placements using Blender.
For the conference room layout~(Figure~\ref{fig_exp_3d_a}), we define three scene versions:
\textit{ConferenceV1} includes both Chairs~1 and~2,
\textit{ConferenceV2} includes only Chair~1, and
\textit{ConferenceV3} includes neither Chair~1 nor Chair~2.
For the office room layout (Figure~\ref{fig_exp_3d_b}), we similarly define three scene versions:
\textit{OfficeV1} includes both Tables~1 and~2,
\textit{OfficeV2} includes only Table~1, and
\textit{OfficeV3} includes neither Table~1 nor Table~2.

In each scene version, a receiver with a fixed antenna array is positioned, and transmitters are randomly distributed throughout the space. 
A total of 4,416, 4,453, 3,107, 8,481, 7,274, and 4,894 transmitters are placed to collect spatial spectra for each scene version using MATLAB ray tracing simulation~\cite{matlab_indoor_simulation}.
Unless stated otherwise, 80\% of the data is for training and 20\% for testing.

\textsc{(iii)}
Additional layouts, \ie a bedroom and an outdoor scene, are described in~§E of the \textit{Supplementary Materials}.

\begin{table}[t]
\centering
\caption{Model performance for the single-scene settings.}
\label{table_overall_single}
\begin{tabular}{L{0.54in}C{0.48in}C{0.48in}C{0.48in}C{0.48in}}
\toprule
Models & MSE$\downarrow$  & LPIPS$\downarrow$  & PSNR$\uparrow$  & SSIM$\uparrow$ \\
\midrule
KNN & 0.089 & 0.357 & 15.16 & 0.543 \\
KNN-DL & 0.048 & 0.198 & 20.81 & 0.675 \\
\nerft & 0.052 & 0.274 & 19.93 & 0.704 \\
\midrule
\rowcolor{gray!20} \ourSystem & 0.038 & 0.136 & 21.94 & 0.766 \\
\bottomrule
\end{tabular}
\end{table}

\noindent
\textbf{\underline{\textbullet~Baselines}}
We compare \ourSystem against three baselines.

\noindent
\textsc{(i)~K-Nearest Neighbors (KNN):}
It predicts the spatial spectrum at a target transmitter location by averaging the spectra of the \( L \) nearest neighbors, ensuring~\( L \) matches~\ourSystem for a fair comparison.

\textsc{(ii)~KNN-DL:}
Similar to KNN, this approach assigns a weight matrix of dimension $(N_a, N_e)$ to each neighbor instead of using equal averaging. 
Each pixel of the spectrum is given a learnable weight, and the target spectrum is obtained by a weighted summation of all neighbors' spectra using their corresponding weight matrices. 
The weight matrices are optimized using an $\ell_2$ loss. 
Further details are provided in~§A.2 of the \textit{Supplementary Materials}.

\textsc{(iii)}~\nerft:
It is introduced in~§\ref{sec_nerf2_intro}. 
NeWRF~\cite{lunewrf} is essentially the same as \nerft, but additionally uses DoA measurements to reduce the number of rays. 
Since DoA data is difficult to obtain in practice, we treat \nerft and NeWRF as equivalent methods in our evaluation. 
Other methods,~\eg~WiNeRT~\cite{orekondy2022winert} and RFScape~\cite{chen2025radio}, are not directly comparable since they require prior scene models.

\noindent
\textbf{\underline{\textbullet~Metrics}}
Mean Squared Error (MSE) measures the average squared difference between the predicted and ground-truth signal power across all directions in the spatial spectrum, serving as the domain loss. 
Since the spatial spectrum (see Figure~\ref{fig_spectrum_b}) can be visualized as an image, image quality metrics are used to assess pixel-level differences and structural consistency~\cite{wang2004image}, capturing directional patterns. 
Accordingly, we evaluate performance using four widely adopted metrics: MSE$\downarrow$, Learned Perceptual Image Patch Similarity (LPIPS$\downarrow$), Peak Signal-to-Noise Ratio (PSNR$\uparrow$), and Structural Similarity Index Measure (SSIM$\uparrow$).

\subsection{Single-Scene Performance}\label{sec_single}

There are seven scenes in total: one from the RFID dataset and six from the MATLAB dataset~(ConferenceV1--V3, OfficeV1--V3). 
In this section, we train and evaluate a separate model for each scene, using 80\% of the data for training and the remaining 20\% for testing.

\begin{figure}[t]
\centering
{\includegraphics[width=.45\textwidth]{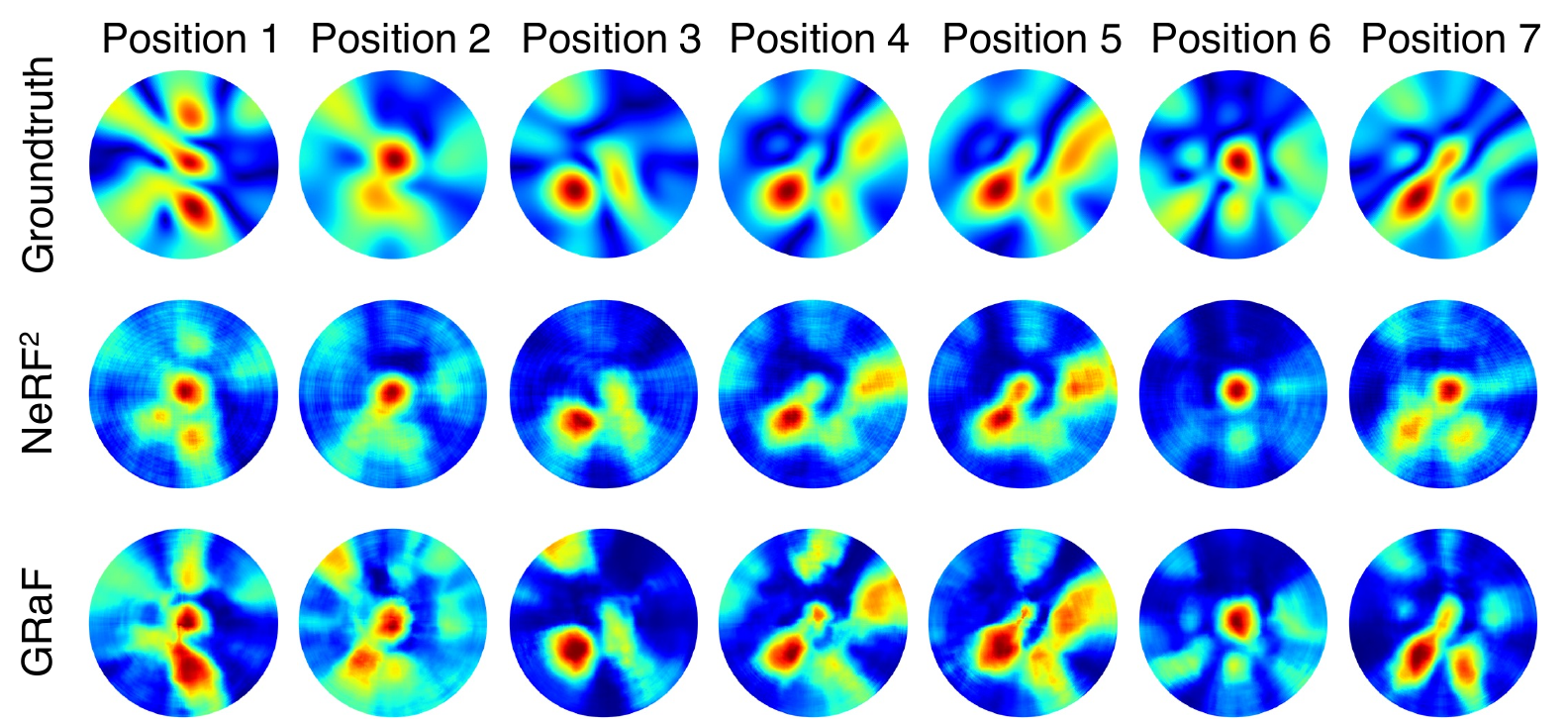}}
	\vspace{2pt}
\caption{Visual comparison of synthesized spatial spectra.}
	\label{fig_vis_d1}
\end{figure}

\noindent
\textbf{Analysis.}  
We first present the spatial spectra of four randomly selected transmitter locations, alongside the spectra synthesized by \nerft and \ourSystem, as illustrated in Figure~\ref{fig_vis_d1}.
Visually, the spectra synthesized by \ourSystem closely match the ground truth, outperforming those generated by \nerft.
Consistent with this visual comparison, \ourSystem achieves superior metric scores, as illustrated in Table~\ref{table_overall_single}. 
Compared to the KNN and KNN-DL, \ourSystem improves PSNR by 44.7\% and 5.5\%, respectively. 
These results demonstrate that neighboring spatial spectra are indeed helpful for predicting the target transmitter's spectrum. 
However, KNN's simple averaging approach fails to capture the intricate spatial relationships among the neighboring spectra. 
KNN-DL further improves upon KNN's performance, indicating that a careful interpretation of neighboring spectra can enhance the prediction accuracy. 
Nevertheless, KNN-DL does not account for the geometric relationships among spectra, limiting its effectiveness.
Finally, \ourSystem outperforms \nerft by 26.9\%, 50.4\%, 10.2\%, and 8.8\% in terms of MSE, LPIPS, PSNR, and SSIM, respectively. 
This improvement highlights \ourSystem's ability to leverage neighboring spatial spectra to learn latent voxel features that capture complex propagation behaviors and dynamically assign weights using an attention mechanism during ray tracing.

\subsection{Generalization to Unseen Scenes}\label{sec_eval_cross}

We evaluate the generalization capability of \ourSystem across unseen scenes, particularly when object configurations within the layout are changed. 
Specifically, we train two models separately for the conference room and office room layouts. 
For the conference room, the model is trained on~\text{ConferenceV1} and tested on the test splits of \text{ConferenceV2} and \text{ConferenceV3}. 
For the office room, the model is trained on \text{OfficeV1} and tested on \text{OfficeV2} and \text{OfficeV3}. 
All baseline methods are trained and evaluated under the same settings for fair comparison. 
The inverse setting,~\ie~the training and testing scenes are swapped, is provided in~§E.2 of the \textit{Supplementary Materials}.

\noindent
\textbf{Analysis.}
Table~\ref{table_overall_cross_s1} presents the average quantitative results for the two layout models.
Compared to the single-scene setting~(Table~\ref{table_overall_single}), KNN achieves similar performance in both scenarios, highlighting that neighboring spatial information consistently provides value. 
In contrast, the performance of~KNN-DL and~\nerft declines in the unseen scene setting due to their reliance on models overfitted to the training scenes, which limits their generalization capabilities.
However, because the differences between the training and testing scenes involve only minor modifications, such as adding or removing one or two tables or chairs, the performance degradation is moderate, with PSNR drops of \(8.51\%\) for KNN-DL and \(12.90\%\) for \nerft.
By comparison, \ourSystem achieves the highest accuracy in both settings by leveraging geometry-aware latent RF radiance fields and a neural ray tracing algorithm, enabling effective generalization across diverse scene layouts.

\begin{table}[t]
\centering
\caption{Model performance on the unseen scenes settings.}
\label{table_overall_cross_s1}
\begin{tabular}{L{0.54in}C{0.48in}C{0.48in}C{0.48in}C{0.48in}}
\toprule
Models & MSE$\downarrow$  & LPIPS$\downarrow$  & PSNR$\uparrow$  & SSIM$\uparrow$ \\
\midrule
KNN & 0.083 & 0.361 & 14.77 & 0.552 \\
KNN-DL & 0.053 & 0.279 & 19.04 & 0.614 \\
\nerft & 0.065 & 0.337 & 17.36 & 0.691 \\
\midrule
\rowcolor{gray!20} \ourSystem & 0.039 & 0.215 & 20.96 & 0.705 \\
\bottomrule
\end{tabular}
\end{table}

\subsection{Generalization to Unseen Layouts}\label{sec_eval_cross_s2}

We further evaluate the generalization capability of \ourSystem under a more challenging setting. 
Specifically, a model is trained on all three versions of the conference layout, \ie the training splits of ConferenceV1–V3, and tested on all three versions of the office layout, \ie OfficeV1–V3.
The reverse experiment is also conducted, where the model is trained on OfficeV1–V3 and tested on ConferenceV1–V3.
For fair comparison, the three baseline methods are evaluated under the same experimental settings.

\noindent
\textbf{Analysis.}  
Table~\ref{table_overall_cross_s2} presents the quantitative results for unseen layouts. 
KNN demonstrates consistent performance across settings with slight PSNR fluctuations, reflecting the effectiveness of neighboring spatial spectrum information. 
In contrast, KNN-DL and \nerft show significant performance declines in unseen layouts compared to single-scene settings. 
KNN-DL's PSNR drops by \(28.2\%\) and \nerft's by \(35.9\%\), highlighting their limited generalization due to scene-specific training. 
\ourSystem achieves the highest accuracy across all settings, demonstrating robust generalization. 
This is attributed to its geometry-aware voxel features and neural ray tracing, which enable it to adapt to diverse scenes. 
However, the spectra synthesized by \ourSystem in cross-layout experiments are of lower quality than those in single-scene experiments. 
This performance decline is due to challenges posed by varied layouts and object materials~\cite{wong1984conductivity}.

The trained \ourSystem is also evaluated on additional layouts~(Bedroom and Outdoor) as well as a real-world dataset, introducing diverse spatial characteristics. 
The Bedroom features confined spaces with rich multipath propagation; 
the Outdoor represents a large-scale open environment with sparse reflectors; 
and the RFID dataset introduces real-world noise, hardware variability, and temporal effects.
Results are provided in~§E.3 of the \textit{Supplementary Materials}.

\begin{table}[t]
\centering
\caption{Model performance on the unseen layouts settings\tnote.}
\label{table_overall_cross_s2}
\begin{threeparttable}
\begin{tabular}{L{0.54in}C{0.48in}C{0.48in}C{0.48in}C{0.48in}}
\toprule
Models & MSE$\downarrow$  & LPIPS$\downarrow$  & PSNR$\uparrow$  & SSIM$\uparrow$ \\
\midrule
KNN & 0.085 & 0.359 & 15.32 & 0.539 \\
KNN-DL & 0.087 & 0.429 & 14.94 & 0.498 \\
\nerft & 0.092 &  0.477 & 12.76 & 0.481 \\
\midrule
\rowcolor{gray!20} \ourSystem & 0.042 & 0.268 & 17.81 & 0.629 \\
\bottomrule
\end{tabular}
\vspace{-0.05in}
\end{threeparttable}
\end{table}

\begin{table}[t]
\centering
\caption{Ablated versions of \ourSystem in unseen scene settings.}
\label{table_ablation_study}
\begin{tabular}{L{1.8in}C{0.48in}C{0.48in}}
\toprule
Model variations     & LPIPS$\downarrow$ & PSNR$\uparrow$   \\ 
\midrule
Without cross-attention layer  & 0.239 & 19.37       \\ 
Without neural ray tracing     & 0.379 & 16.79       \\ 
Full model (\ourSystem)        & 0.215 & 20.96       \\ 
\bottomrule
\end{tabular}
\end{table}

\subsection{Ablation Study}\label{sec_eval_ablation_study}

This section follows the unseen-scene setting described in §\ref{sec_eval_cross}, with results shown in Table~\ref{table_ablation_study}. 
The impact of neighbor number~$L$ is given in §E.4 of the \textit{Supplementary Materials}.

\noindent
\textbf{Latent RF Radiance Field.}
In the parameterized function~\(\mathcal{T}_\Psi\) of our model, we adopt a geometry-aware cross-attention layer to learn contextual latent representations. 
To evaluate the effect of attention mechanisms, we replace the cross-attention layer with a simpler dot-product attention layer. 
A comparison between the first and third rows in Table~\ref{table_ablation_study} shows that \ourSystem achieves slightly better performance with the cross-attention mechanism. 
This result suggests that \ourSystem’s generalization ability stems from its capacity to model interactions across neighboring spatial spectra.

\noindent
\textbf{Neural Ray Tracing.}  
We replace our neural ray tracing algorithm with a simplified method based on Equation~(\ref{eqn_tracing_rf}), which models each voxel using only two attributes: signal emission and attenuation. 
A comparison between the second and third rows in Table~\ref{table_ablation_study} shows that our neural approach outperforms the simplified method, highlighting the advantages of voxel-level vector representations and the model’s ability to learn adaptive fusion weights.
The difference between the ablated version of~\ourSystem~(``without neural ray tracing'' in Table~\ref{table_ablation_study}) and \nerft lies in scene representation. 
While the ablated \ourSystem retains the parameterized function~\(\mathcal{T}_\Psi\) for processing neighboring spectra, it simplifies voxel modeling and uses the ray tracing module from \nerft.
As shown in Table~\ref{table_overall_cross_s1}, \nerft achieves a PSNR of 17.36~dB, while the ablated version of \ourSystem yields 16.79~dB~(Table~\ref{table_ablation_study}).
This performance gap underscores the importance of \ourSystem’s full pipeline.

\begin{figure}[t]
\centering

\begin{subfigure}{.45\linewidth}
  \centering
  \includegraphics[width=\linewidth]{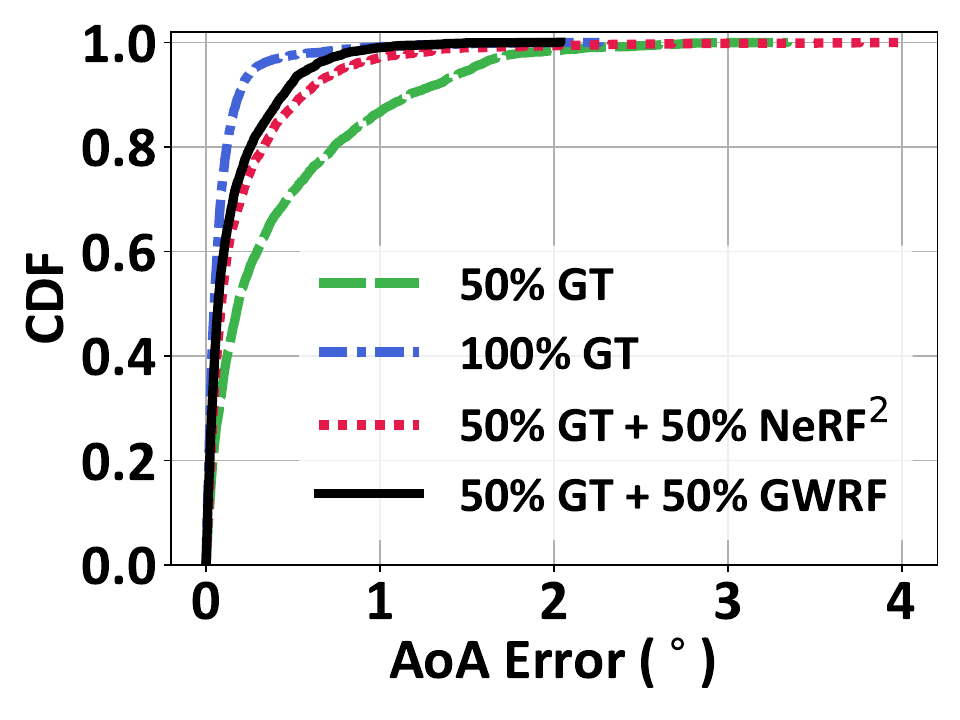}
  \caption{For AANN training}
  \label{fig_case_study_a}
\end{subfigure}
\hfill
\begin{subfigure}{.45\linewidth}
  \centering
  \includegraphics[width=\linewidth]{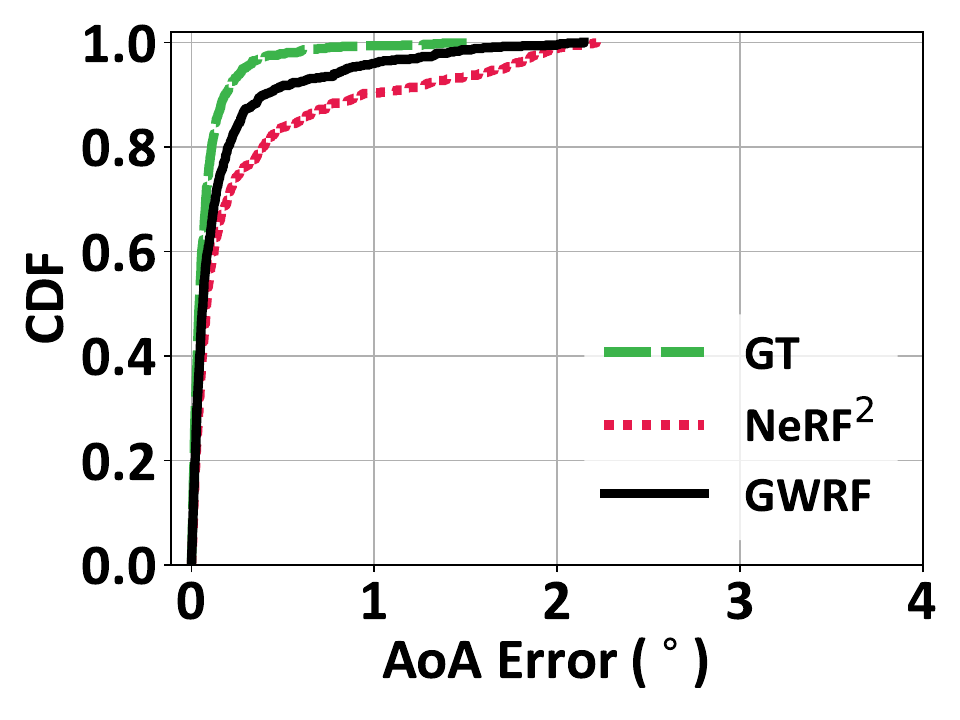}
  \caption{For AANN testing}
  \label{fig_case_study_b}
\end{subfigure}

\caption{Spectrum-based AoA estimation via AANN~\cite{an2020general}.}
\label{fig_case_study}
\vspace{-0.05in}
\end{figure}

\subsection{Impact of RF Frequency Bands}\label{sec_parameter_study}

To examine the impact of frequency bands, we collect three datasets in a conference room~(Figure~\ref{fig_exp_3d_a}) at 928\,MHz, 2.412\,GHz, and 5.805\,GHz. 
\ourSystem is trained on each dataset. 
As shown in Table~\ref{table_para_material}, \ourSystem adapts well to different bands while maintaining high-quality synthesized spectra. 
This design is justified since each band supports different wireless technologies~(\eg 5G, WiFi), requiring dedicated hardware for signal reception and processing~\cite{song2025terahertz, wesling2021heterogeneous, drubin2023four}.

Additionally, in Table~\ref{table_rf_detailed}, we evaluate a version of \ourSystem trained only on the 2.412\,GHz dataset~(voxel size~\(\lambda \approx 0.124\,\text{m}\)) and directly tested on the 928\,MHz and 5.805\,GHz datasets without retraining. 
During testing, the voxel size is held fixed at \(0.124\,\text{m}\).
This cross-frequency evaluation reveals a noticeable performance drop, primarily due to two factors. 
First, RF propagation characteristics vary with frequency. 
At~928\,MHz, signals experience stronger penetration and reduced attenuation, while at~5.805\,GHz, they are more prone to scattering and absorption. 
These variations in wave behavior limit the model’s ability to generalize across bands. 
Second, maintaining a fixed voxel size leads to spatial resolution mismatch. 
For 928\,MHz~(\(\lambda \approx 0.323\,\text{m}\)), the voxel grid is unnecessarily fine, preserving details but increasing computational cost. 
In contrast, for~5.805\,GHz~(\(\lambda \approx 0.052\,\text{m}\)), the voxel size is too coarse to resolve fine-scale propagation effects.

\subsection{Case Study: Angle of Arrival~(AoA) Estimation}\label{sec_eval_case_study}

The synthesized spatial spectrum can be used for downstream tasks such as Angle of Arrival~(AoA) estimation for wireless localization~\cite{an2020general}. 
We use an angular artificial neural network~(AANN), which consists of a ResNet-50 backbone followed by an MLP head, to process the spectrum and estimate the~AoA. 
This estimated AoA corresponds to the line-of-sight~(LoS) propagation direction between the transmitter and receiver, enabling transmitter localization.
By generating synthetic training datasets, \ourSystem significantly reduces the data collection effort required to train the AANN. 
To demonstrate the benefits of the high-quality spectra synthesized by \ourSystem, we evaluate using the following two strategies for the training and testing stages:

\begin{table}[t]
\centering
\caption{Training separately on each frequency band.}
\label{table_para_material}
\begin{tabular}{L{0.65in}C{0.62in}C{0.62in}C{0.62in}}
\toprule
& 928\,MHz & 2.412\,GHz & 5.805\,GHz \\
\midrule
PSNR$\uparrow$ & 25.70 & 24.53 & 24.91 \\
\bottomrule
\end{tabular}
\vspace{-0.1in}
\end{table}

\begin{table}[t]
\centering
\caption{Training only on the 2.412\,GHz frequency band.}
\label{table_rf_detailed}
\begin{tabular}{L{0.54in}C{0.48in}C{0.48in}C{0.48in}C{0.48in}}
\toprule
 & MSE$\downarrow$ & LPIPS$\downarrow$ & PSNR$\uparrow$ & SSIM$\uparrow$ \\
\midrule
928\,MHz   & 0.104 & 0.362 & 14.84 & 0.576 \\
5.805\,GHz & 0.218 & 0.467 & 11.61 & 0.493 \\
\bottomrule
\end{tabular}
\end{table}

\noindent
\textbf{(i) Synthesized Spectra for AANN Training.}  
An AANN is trained on four datasets: (1) 50\% ground truth (GT), (2) 50\% GT combined with 50\% \nerft-synthesized spectra, (3) 50\% GT combined with 50\% \ourSystem-synthesized spectra, and (4) 100\% GT. 
All training datasets use the same transmitter locations, and each trained AANN is evaluated on a shared testing set with consistent transmitter positions. 
Figure~\ref{fig_case_study_a} presents the cumulative distribution function (CDF) of the estimated AoA error. 
The \ourSystem-augmented dataset reduces the estimation error by 61.6\% compared to training on 50\% GT alone, and by 25.8\% relative to the \nerft-augmented dataset. 
These results demonstrate that \ourSystem produces high-quality spectra that effectively improve localization model training.

\noindent
\textbf{(ii)~Synthesized Spectra for AANN Testing.}  
An AANN is trained on GT spectra and evaluated on three types of test datasets: GT, \nerft-synthesized spectra, and \ourSystem-synthesized spectra. 
All test spectra correspond to the same transmitter locations. 
Figure~\ref{fig_case_study_b} demonstrates that \ourSystem's spectra yield more accurate AoA estimations, highlighting the superior fidelity of \ourSystem's synthesized spectra.

%% file: tex/6_Conclusion.tex
\section{Conclusion}\label{sec_conclusion}

This paper presents \ourSystem, a generalizable RF radiance field framework for spatial spectrum synthesis. 
We first establish the interpolation theory in the RF domain as the theoretical foundation. 
\ourSystem integrates a geometry-aware latent RF radiance field to capture spatially correlated propagation characteristics and employs a neural ray tracing algorithm to aggregate these features for accurate spectrum computation. 
Experimental results show that \ourSystem generalizes across scenes and enhances downstream task performance.

\section*{Acknowledgements}

This research was funded in part by the National Science Foundation through grants CNS-2239458 and CCSS-2525613. 
Any opinions, findings, and conclusions expressed in this material are those of the authors and do not necessarily reflect the views of the funding agencies.